\def\year{2021}\relax
\documentclass[letterpaper]{article} % DO NOT CHANGE THIS
\usepackage{aaai22}  % DO NOT CHANGE THIS
\usepackage{times}  % DO NOT CHANGE THIS
\usepackage{helvet} % DO NOT CHANGE THIS
\usepackage{courier}  % DO NOT CHANGE THIS
\usepackage[hyphens]{url}  % DO NOT CHANGE THIS
\usepackage{graphicx} % DO NOT CHANGE THIS
\usepackage{natbib}  % DO NOT CHANGE THIS AND DO NOT ADD ANY OPTIONS TO IT
\usepackage{caption} % DO NOT CHANGE THIS AND DO NOT ADD ANY OPTIONS TO IT
\usepackage{booktabs}
\usepackage{xcolor}

\title{\Huge{Detecting Anti-Vaccine Users on Twitter}}
\author {
    Matheus Schmitz,
    Goran Muric, 
    Keith Burghardt  \\
}
\affiliations {
    USC Information Sciences Institute \\
    \{mschmitz, gmuric, keithab\}@isi.edu
}

\begin{document}
\maketitle

\begin{abstract}
Vaccine hesitancy, which has recently been driven by online narratives, significantly degrades the efficacy of vaccination strategies, such as those for COVID-19. Despite broad agreement in the medical community about the safety and efficacy of available vaccines, a large number of social media users continue to be inundated with false information about vaccines and are indecisive or unwilling to be vaccinated. The goal of this study is to better understand anti-vaccine sentiment by developing a system capable of automatically identifying the users responsible for spreading anti-vaccine narratives. We introduce a publicly available Python package capable of analyzing Twitter profiles to assess how likely that profile is to share anti-vaccine sentiment in the future. The software package is built using text embedding methods, neural networks, and automated dataset generation and is trained on several million tweets. We find this model can accurately detect anti-vaccine users up to a year before they tweet anti-vaccine hashtags or keywords. We also show examples of how text analysis helps us understand anti-vaccine discussions by detecting moral and emotional differences between anti-vaccine spreaders on Twitter and regular users. Our results will help researchers and policy-makers understand how users become anti-vaccine and what they discuss on Twitter. Policy-makers can utilize this information for better targeted campaigns that debunk harmful anti-vaccination myths. 
\end{abstract}

\section{Introduction}
Anti-science, and especially anti-vaccine, attitudes are present within a large and recently active minority \cite{Germani2020, Murphy2021}.  Anti-vaccine protesters are partly responsible for a significant resurgence of measles and other diseases for which vaccines have existed for decades \cite{Smith2017}. Vaccine hesitancy is especially problematic for COVID-19, which continues to be an epidemic, especially within the United States, due in part to individuals not socially distancing, wearing masks, and becoming vaccinated, despite the advice of the medical community. The rapid spread of anti-science conspiracy theories and polarization online is one reason behind these attitudes \cite{Druckman2020, Rao2020}. This motivates our research in applying machine learning to \emph{predict}, rather than just classify, if individuals will become anti-vaccine. Such a tool can help policy-makers and researchers uncover factors that draw people to become anti-vaccine and design targeted campaigns that reduce vaccine hesitancy. We would also like to do this at scale by applying these predictions to social media users. 

%We therefore ask whether social media text can be leveraged to predict anti-vaccine sentiment, specifically if users share anti-vaccine narritives.

In this paper, we create an algorithm we call \textit{AVAXTAR} that evaluates the likelihood a Twitter account will spread anti-vaccine narratives. This code is freely available as a Python package: https://bit.ly/3uGXep8. The underlying model in AVAXTAR is trained to predict whether a Twitter account will spread anti-vaccine hashtags or keywords \emph{up to a year in advance}. A researcher can provide a Twitter ID or unique screen name to AVAXTAR and the package then retrieves the target account's recent activity. From this activity, the package returns the account's probability of future anti-vaccine discussion. Twitter is explored in this paper because it is a popular social media website with an ongoing problem of anti-science rhetoric \cite{Rao2020}. 

We also leverage the dataset gathered for model training to explore the textual differences between the tweets posted by users who do and users who do not spread the anti-vaccine narratives. This analysis provides clues about the underlying reasons for anti-vaccine sentiment as well as the rhetorical devices people use to spread misinformation. Overall, the presented work provides a new method to understand the recent uptick in anti-vaccine sentiment by identifying users prone to disseminating such messages, and can help policy-makers devise targeted information campaigns. 

%This work also helps researchers and policy-makers to devise information campaigns targeted at vaccine-hesitant individuals.

\section{Related Work}

\subsection{Detecting Vaccine Sentiment}

Several papers on detecting vaccine sentiment have been developed. Wang et al.~(\citeyear{Wang2020}) developed a multi-modal system to classify tweets as containing anti-vaccine sentiment, and the authors achieve a 97\% accuracy and F1-score. A complimentary ablation study displays which sections of images, texts and hashtags were most associated with the label by the model. An alternative data source is presented by Carrieri et al.~(\citeyear{carrieri2021predicting}), who use area-level indicators from Italy to develop various models which both detect vaccine hesitancy by region as well as provide a list of the most important features associated with vaccine hesitancy. A third approach to data collection was employed by Lincoln et al.~(\citeyear{lincoln2022taking}), who collected survey data from adults in five countries, assessing stances on vaccine hesitancy as well a various demographic and psychographic factors. That data is then used to train a random forest classifier, which obtains a recall score of 82\% and precision of 79-82\% depending on country.

Predicting anti-vaccine hesitancy, rather than simply detecting it, is a more difficult problem with comparatively less research. Huang et al.~(\citeyear{huang2017examining}) applied a range of conventional classifiers to detect whether someone received, or intended to receive, an influenza vaccine from a Twitter dataset on tweets about influenza from 2013 to 2016 (three flu seasons). They report F1 scores of up to 82\%. Krishnan et al. (\citeyear{krishnan2021predicting}) applied the same dataset to predict vaccine hesitancy using Latent Dirichlet Allocation and Particle Swarm Optimization and found an F1 score of 84\%. 

\subsection{Text Analysis of Vaccine Hesitancy}
%Most related to the textual analysis presented in this paper is prior work by 
One of the first papers exploring text associated with vaccine hesitancy was by Amin et al.~(\citeyear{amin2017association}) who explored moral values and vaccine hesitancy. They find that harm and fairness foundations are not significantly associated with vaccine hesitancy, but purity and liberty foundations are. Medium-hesitancy adults were twice as likely as low-hesitancy ones to highly emphasize purity, while high-hesitancy adults were twice as likely to strongly emphasize both purity and liberty. A complimentary explanation of the drivers of anti-vaccination behavior comes from Kadam (\citeyear{kadam2017understanding}), who attempts to determine the events and incidences responsible for amplifying pro-vaccination and anti-vaccination sentiments. The author reports two sets of hashtags which are associated with positive and negative sentiments on vaccines. %The importance of Twitter in vaccination-related behavior is further evidenced by 
Germani et al.~(\citeyear{germani2021anti}), meanwhile, explore data on anti-vaccination supporters within Twitter, identifying they share more conspiracy theories, make larger use of emotional language, are more engaged in discussions and share their contents from a poll of strong influencers. The authors observe that the anti-vaccine movement's success depends on a strong sense of community, reliant on the content produced by a minority of users, with the larger community working as an amplifier bringing anti-vaccination discourse the platform at large. 

Felmlee et al.~(\citeyear{Felmlee2020}) exploit a Twitter policy change related to abusive content to test the effectiveness of organizational policies aimed at stemming online harassment. They find evidence of a modest positive shift in the sentiment of tweets with slurs targeting women and/or African Americans. Retweeted messages are more negative than those not forwarded. These patterns suggest that organizational ``anti-abuse'' policies can play a role in stemming hateful speech on social media without inflaming further abuse. Network effects and the out-sized impact of certain users can be appreciated in Radzikowski el al.~(\citeyear{radzikowski2016measles}), who explores Twitter narratives regarding vaccination in the aftermath of the 2015 measles outbreak. They find stories contributed by news organizations have a higher impact compared to direct tweets by health organizations in communicating health-related information. A complimentary example of low relevance factors comes by Hornsey et al.~(\citeyear{hornsey2020vaccine}), who study the relationship between trust in Complementary and Alternative Medicines (CAM) and vaccine hesitancy, finding that trust in CAM is only a weak predictor of vaccine hesitancy.

\subsection{Community Analysis}
Several papers have explored relationships between the social network and vaccine hesitancy. Francia et al.~(\citeyear{francia2019social}) combines both community detection and text analytics methodologies to characterize the Twitter debate on vaccination in Italy. The authors find a strong association between political leaning and vaccination stance, a high similarity between groups opposing vaccines entirely and those opposing vaccine mandates, and large passive communities with a focus on non-vaccine topics but with a pro- or anti-vaccine stance. Bello et al.~(\citeyear{bello2017detecting}) create a graph of Twitter discussion communities on vaccination. Adding geolocation information the authors generate a summary of the relevance of vaccination topics across countries and tag communities to their associated countries, with the US hosting the majority of the anti-vaccination movement. Combing Machine Learning and graph models, Yuan et al.~(\citeyear{Yuan2019}) explore tweets related to the MMR vaccine published after the 2015 California Disneyland measles outbreak. They use machine learning to classify users into anti-vaccination, neutral, and pro-vaccination groups. Using community detection, the authors show that pro- and anti-vaccine users share predominantly in-group narratives. Moreover, anti-vaccine communities are highly clustered and enclosed. %This work contrasts with our analysis based on state-of-the-art modeling of COVID-19 discussions in which we can predict opinions up to a year in advance and apply NLP methods to understand the emotions, morals, and common text of anti-vaccine users. 
Schmidt et al.~(\citeyear{SCHMIDT20183606}) detect the emergence of communities on Facebook (rather than Twitter, as in previous work), showing the consumption of content about vaccines is dominated by the echo chamber effect and that polarization has increased over the years. Well-segregated communities emerge from the users’ consumption habits with few cross-ideological content consumption.

Kang et al.~(\citeyear{KANG20173621}) constructed semantic networks of vaccine articles shared by Twitter users, finding negative vaccine-sentiment networks centered on larger organizations, while positive-sentiment networks show more cohesive discourse, with discussions about parents, vaccines, and their non-association with autism. Analysis by Featherstone et al.~(\citeyear{FEATHERSTONE2020100105}) discuss analysis of tweets about childhood vaccines, finding a wide sharing of vaccine misinformation within a well-connected anti-vaccine community, with a few influential users located in certain geo-located clusters producing the bulk of the content. The authors also find that pro-vaccine and anti-vaccine tweets are predominantly of negative tone, although such negativity can also be a reflection of the incentives created by social media feed algorithms or negativity bias \cite{Peeters1990}. A theoretical take on communities is presented by Barlett et al.~(\citeyear{Barlett2018}), who discuss how social media anonymity can increase cyberbullying perpetration. Further prescriptive guidance is provided in Wilson et al.~(\citeyear{Wilson2014}), in the form of a framework for using mobile technology to increase vaccine confidence as well as to create a surveillance and response system that monitors digital conversations on the topic and provide public health officials early warning about clusters of people with fading confidence in vaccination.

\subsection{Contributions of our Research}
Our work contrasts with these previous papers by developing and distributing an open-source library to empower future research in the field. Many previous papers developed bespoke machine learning models as steps to achieve other goals, and few have been applied to classify anti-vaccine sentiment. More specifically, this work is unique in predicting the vaccination stance of users up to one year before they begin to tweet anti-vaccine rhetoric. In addition, our comparative analysis of transformer-based classifiers within COVID-19 tweet discussions improves the robustness of our methods. Finally, we employ natural language processing tools to understand the differences in emotions, morals, and common words of anti-vaccine users at scale.

\section{Methods}

\subsection{Data Collection}
The \textit{AVAXTAR} classifier is trained on a comprehensive labeled dataset of tens of millions of tweets from approximately 130 thousand Twitter accounts. Each account from the dataset was assigned one of two labels: \textit{1} for the accounts that actively spread anti-vaccination hashtags or keywords ($\approx 70$ thousand) and \textit{0} for the accounts that do not tweet anti-vaccine hashtags or keywords ($\approx 60$ thousand). By leveraging Twitter's Academic Research Product Track, we were able to access the full archival search and overcome the limit of 3,200 historical tweets of the standard API. We therefore collect almost all historical tweets of most queried accounts (for a small fraction of accounts that are highly active we interrupted the collection prematurely, due to Twitter's API limitations). Sample tweets from users belonging to each class are shown in Table~\ref{tab:avax_example}.

\begin{table*}[h]
    \centering
    \caption{Sample tweets from each class}
    \label{tab:avax_example}
    \scriptsize
    \begin{tabular}{p{0.08\linewidth} p{0.4\linewidth} p{0.4\linewidth}} 
    \toprule
     & \textbf{Account 1} & \textbf{Account 2}\\
    \midrule
    Tweet 1 & As first runner-up to my esteemed @StarTrek colleague @levarburton [when we appeared on \#TheWeakestLink ], I would be honored to try my hand as \#Jeopardy guest host. My experience as a science presenter for @exploreplanets emboldens me to \#boldlygo !
            & Even with the inflated (for scaremongering purposes I can only assume) figure of 126k people who died WITH (not OF remember) covid19, that would mean 
                that in a whole year this "killer virus" hasnt even managed to kill 0.19\% of almost 68 million people in the UK. "pandemic" \\ [0.3cm]
    Tweet 2 & @SpaceX is daring some mighty things. To the stars!
            & \#NoVaccinePassportsAnywhere \#NoVaccinePassportAnywhere \#NoVaccinePassports \#NoVaccinePassport \#novaccinatingthechildren \\ [0.3cm]
    Tweet 3 & Congratulations to all at Blue Origin. Nicely done!
            & It's a new week, and no better a time to remind @nadhimzahawi that he's a disgusting, two-faced parasite whose name will forever be synonymous with lies, corruption and bloodshed. Please help him get the message \\ [0.3cm]
    Tweet 4 & We visited Virgin Galactic back in 2018. Flew the simulator. Looked like it was going to fly well. And it did. Congratulations to All!
            & \#NoVaccinePassportsAnywhere  \#NoVaccinePassports \#MedicalApartheid \#wedonotconsent Really handy website to contact your MP directly...\\ [0.2cm]
    \midrule 
    \textbf{Anti-Vaccine Probability} & \textbf{0.0705813} & \textbf{0.99880254}\\
    \bottomrule
    \end{tabular}
\end{table*}

\textbf{Collecting anti-vaccine samples.} In this study, we label anti-vaccination users as ``1'', where ground-truth anti-vaccine users come from an existing dataset of anti-vaccine Twitter accounts and their respective tweets, collected and published by Muric et al.~(\citeyear{Muric2021}). The authors first used a snowball method to identify a set of hashtags and keywords associated with the anti-vaccination movement, and then queried the Twitter API and collected the historical tweets of accounts that used any of the identified hashtags or keywords. More than 135 million tweets were collected from more than 70 thousand accounts.

\textbf{Collecting not anti-vaccine samples.} To collect data on examples of tweets and users that are not anti-vaccine (labeled ``0''), we first performed a similar approach to Muric et al.~(\citeyear{Muric2021}), and queried the Twitter API to get historical tweets of accounts that do not use any of the predefined keywords and hashtags. Using this method, we collected the tweets of $\sim 30$ thousand accounts that do not spread anti-vaccination narratives or are impartial about the topic. This sample most likely represents typical Twitter users. We then enrich this sample by gathering the tweets from accounts that are likely proponents of the vaccination. We identify the proponents of the vaccines in the following way: First, we identify the set of twenty most prominent doctors and health experts active on Twitter. Then, we manually collected the URLs of Lists\footnote{Twitter Lists allow users to customize, organize and prioritize the tweets they see in their timelines. Users can choose to join Lists created by others on Twitter.} those health experts they made on Twitter. We specifically searched for lists with epidemiology-related names (e.g., ''coronavirus experts'' or ''epidemiologists''). From those lists, we collected approximately one thousand Twitter handles of prominent experts and doctors who tweet about the coronavirus and the pandemic. We went through their latest 200 tweets and collected the Twitter handles of users who retweeted their tweets. That became our pool of pro-vaccine users. The users who retweeted many distinct experts were more likely to be included than users who retweeted a few. Finally, we collected the historical tweets of users from the pro-vaccine pool. This way we collected more than 50 million tweets from more than 30 thousand accounts that are most likely pro-vaccine, therefore $60$ thousand accounts and more than 100 million tweets are gathered from users who were not anti-vaccine.

\subsection{Classification System}
\textbf{Generating Training Dataset.}  If we na{\"i}vely train on these labeled data, we might only capture whether users did or did not use particular keywords and hashtags, which would also artificially inflate the model accuracy. We instead want to capture more nuanced user language, which allows for a more robust and generalizable model. To address this, we train on anti-vaccine accounts before its \textit{labeling date}, the first date in which the account published a tweet that contained one of the predefined anti-vaccination hashtags and keywords defined in Muric et al.~(\citeyear{Muric2021}). For the not anti-vaccine user cohort, their labeling date was the date of their most recent tweet. All tweets from the 15 months prior to that date were considered in model training, with samples being created within a 90-day time windows prior to the labeling date: [0-90), [60-150), [120-210), [180-270), [240-330), [300-390), [360-450) days. For each time window, all tweets from a given user were merged into a single document. The resulting training dataset contains $~130$ thousand users, 70\% of which were sampled for training, 15\% for validation and 15\% for testing.

\textbf{Model Selection.} We consider three candidate sentence embedding models which are state-of-the-art in this category while also being sufficiently small that users can run them on regular desktop machines, as our goal is to provide an accessible and useful package. The models tested were: Sent2Vec~\cite{Gupta19}, Sentence-MPNet~\cite{Kaitao2020}, and Sentence-Distill-Roberta~\cite{Liu2019}.
Techniques for adapting word-level transformers into sentence-level models come from Reimers et al. (\citeyear{reimers-2019-sentence-bert}), while distillation is a technique pioneered by Hinton et al. (\citeyear{Hinton2015}) that enables model shrinkage with little or no loss in performance.

Sentence-Distill-Roberta was trained on OpenWebText~\cite{Radford2019}. Sentence-MPNet was trained on Wikipedia and BooksCorpus~\cite{Zhu2015}, OpenWebText~\cite{Radford2019}, CC-News~\cite{Liu2019} and Stories~\cite{Trinh18}. Initially a Sent2Vec embedder pre-trained on Twitter bigrams was employed, but it presented engineering challenges with regards to publishing a package, as the model needs 22GB of working memory, making it too large for most users' computers. We ran experiments with alternative smaller embeddings models, trained on either Wikipedia or the BookCorpus dataset~\cite{Zhu2015}, and we found that a model based on a Wikipedia embbeder presented only a 1\% loss in F1-Score, while being significantly smaller, at 8GB. All comparisons in this paper are made using this Wikipedia-based version of Sent2Vec.

Embeddings are created for all tweets within each user-window. They are then used to train the feed-forward neural network. After fine tuning the architecture and hyper-parameters, the final obtained neural network consists of three layers: (1) Fully connected layer of size equals to each models' embedding dimension, (2) Fully connected layer with half the size of the previous layer and (3) Fully connected layer with half the size of the previous layer. In between layers a 40\% dropout rate was applied. We used hyperbolic tangent activation between the layers and a softmax activation to generate prediction confidences. The batch size was 128, binary cross-entropy is used as loss function, and the optimizer is Adaptive Moment Estimation with Weight Decay (AdamW)~\cite{AdamW}.

\textbf{Feature Engineering.} The Twitter API provides a standard output containing a variety of data and metadata for each tweet. Thus, many more potentially useful tweet features are obtained, which are then used to generate several engineered features in an attempt to improve the predictive model. To construct engineered features, we considered factors such as the count and share of tweets, retweets, replies and quotes; the median number of favorites, retweets, replies and quotes that a user's publications receive; the number of days in which the user made a publication; whether the user's account is verified; the average sentiment (positive or negative) of the users posts, obtained with the python wrapper (vaderSentiment) for VADER \cite{Hutto2014}; the number and percentage of total tweets from a user which are retweets of prominent anti-vaccination users; and lastly, the number of times a user shared an URL to websites considered ``Conspiracy Pseudoscience,'' ``Questionable Sources,'' or ``Pro Science'' according to Media Bias/Fact Check (mediabiasfactcheck.com/), a website that rates media outlets on their factual accuracy and political leaning. These features were generated using the same sliding window procedure described above: [0-90), [60-150), [120-210), [180-270), [240-330), [300-390), [360-450) days. The model was then trained with embeddings plus engineered features, embeddings only, and engineered features only, and performance analysis revealed all engineered features to have negligible impact on accuracy, F1-Score, ROC-AUC, and PRC-AUC when used alongside the transformer embeddings. Based on those results, the engineered features were dropped from the model, and the final model thus utilizes only textual embeddings.

\begin{table}[h]
\centering
\caption{Classifier evaluation scores on the test set}
\label{tab:metrics}
\small
\begin{tabular}{r c c c} 
 \toprule
 \textbf{Metric} & \textbf{Sent2Vec} & \textbf{S-MPNet} & \textbf{S-Distill-Roberta}  \\
 \midrule
 Accuracy   & \textbf{0.8566}  & 0.8336  & 0.8350  \\ 
 ROC-AUC    & \textbf{0.9084}  & 0.8821  & 0.8826  \\
 PRC-AUC    & \textbf{0.9601}  & 0.9474  & 0.9473  \\
 Precision  & \textbf{0.8541}  & 0.8317  & 0.8319  \\
 Recall     & \textbf{0.8566}  & 0.8336  & 0.8350  \\
 F1         & \textbf{0.8550}  & 0.8325  & 0.8331  \\
\bottomrule
\end{tabular}
\end{table}

\begin{figure}[h] 
  \centering
  \includegraphics[width=.95\linewidth]{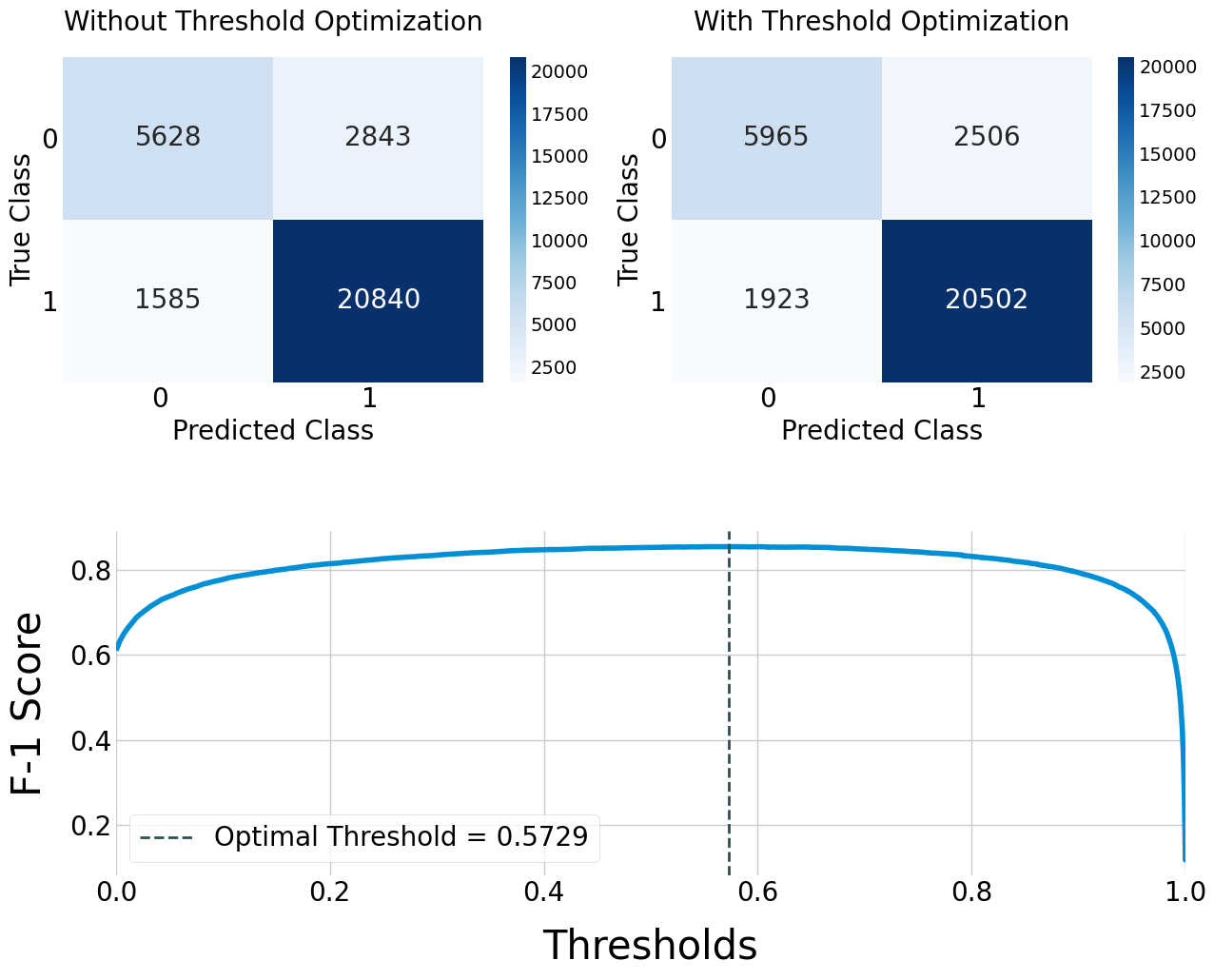}
  \caption{Upper: Confusion Matrix before and after threshold optimization; Lower: Relation between  classification threshold and F1-Score. Optimal threshold for the highest F1-Score is 0.5729.}
  \label{fig:confusion-matrix}
\end{figure}

\textbf{Fine Tuning.} For all models we fine tune the classification threshold to be used, based on maximizing F1 score on the validation set. Using the optimized threshold, the resulting models were then evaluated on a test set of users, achieving the scores shown in table~\ref{tab:metrics}. Based on this analysis, Sent2Vec was chosen as the model to underpin the AVAXTAR package, as it shows the highest performance across all metrics considered. For the chosen model, a test-set Confusion Matrix and F1-Score analysis is shown in Figure~\ref{fig:confusion-matrix}. We find that for the Sent2Vec model a threshold of 0.5729 results in the highest F1 Score, and thus recommend that threshold instead of the more typical threshold of 0.5. 

We assess the model against the full test data (using all windows) as well as against each specific window. As we increase the gap between data collection and the posting period being predicted, from 0 days up to 360 days, we see a slight decrease in performance, as the model must rely only on tweets further in the past. When comparing the one-year gap with zero-days gap, we observe a drop in ROC-AUC from 0.94 to 0.85, as shown in table \ref{tab:test-windows}. We also notice a sparser set of data as we move further back because users are less likely to have continuously tweeted over one year. These results suggests that predicting anti-vaccine sentiment, defined as using anti-vaccine hashtags or keywords, is feasible \emph{up to a year before a user makes any anti-vaccine tweets}.

\begin{table*}[h]
\centering
\caption{Classifier evaluation scores on the test set}
\label{tab:test-windows}
\small
\begin{tabular}{c c c c c c c} 
 \toprule
 \textbf{Test Data Gap} & \textbf{Not Anti-vaccine Samples} & \textbf{Anti-vaccine Samples} & \textbf{Accuracy} & \textbf{F1-Score} & \textbf{ROC-AUC} & \textbf{PRC-AUC} \\
 \midrule
  All Windows  & 8471  & 22425  &  0.8566  &  0.8550  &  0.9084  &  0.9601 \\
  \hline
 $[$0-90)       & 1540  &  10233  &  0.9240  &  0.9216  &  0.9411  &  0.9893 \\ 
 $[$60-150)     & 1427  &   3601  &  0.8566  &  0.8551  &  0.9113  &  0.9596  \\ 
 $[$120-210)    & 1319  &   2495  &  0.8296  &  0.8298  &  0.8901  &  0.9319  \\ 
 $[$180-270)    & 1218  &   1979  &  0.8051  &  0.8048  &  0.8708  &  0.9097 \\ 
 $[$240-330)    & 1115  &   1675  &  0.8047  &  0.8040  &  0.8697  &  0.9021 \\ 
 $[$300-390)    &  998  &   1468  &  0.8049  &  0.8024  &  0.8701  &  0.8987 \\ 
 $[$360-450)    &  854  &   1351  &  0.7927  &  0.7915  &  0.8536  &  0.8941 \\ 
\bottomrule
\end{tabular}
\end{table*}

\textbf{Python Package.} The trained neural network was bundled alongside a script that automates the fetching of the relevant data from the Twitter API. The code was then packaged alongside auxiliary scripts and published to GitHub under the acronym AVAXTAR: Anti-VAXx Tweet AnalyzeR 1.0 and is accessible on GitHub: {https://bit.ly/3uGXep8}. %{https://github.com/Matheus-Schmitz/avaxtar}.
The package abstracts all the feature generation and data manipulation aspects of the task, requiring the user to enter only their Twitter credentials (required for fetching data), alongside a target user's screen name or user id. The provided output consists of a set of probabilities for the user belonging to the ``not anti-vaccine'' class (0) and to the ``anti-vaccine'' class (1).

\section{Data Analysis}

\begin{figure}[h] 
  \centering
  \includegraphics[width=.75\linewidth]{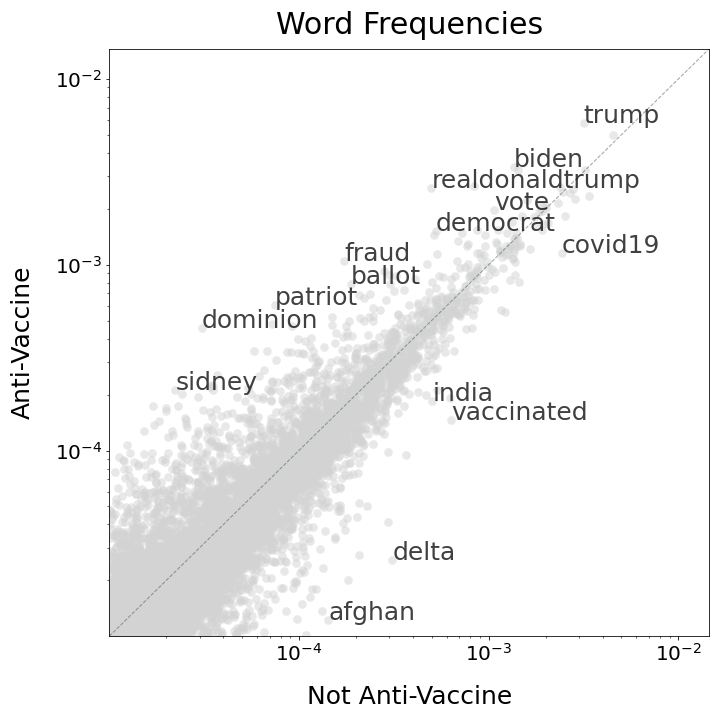}
  \caption{Most frequently used words for anti-vaccine and non anti-vaccine users. A few words used uncommonly often by each respective group are highlighted.}
  \label{fig:frequency}
\end{figure}

A model based on sentence embedding does not necessarily provide insights into the differences between anti-vaccine users and all others. To understand what sets these users apart, we analyze the differences in their text. We first analyze the relative popularity of words used by members of each group, which is shown in Figure~\ref{fig:frequency}; axes are in log-scale and we plot the most common words in each class. We find that, at least among the highest frequency words, the main topic of discourse among Anti-Vaccine users is not vaccination itself, but rather politics in general, with both Trump and Biden as well as ``democrat,'' ``fraud,'' and ``patriot'' among the words whose usage skews the most towards the Anti-Vaccine group. The linguistic differences between Anti-Vaccine users and not Anti-Vaccine users is consistent with research on conspiracy theory content, which found that conspiracy theorists consistently use words like ``stealing'' and ``government'' more often than most users \cite{Klein2019}. The Not Anti-Vaccine users, on the other hand, use COVID-19 and vaccination-related words among its most frequently used words. This is possibly a result of how the Not Anti-Vaccine cohort is defined, with half of its samples being random Twitter users, and the other half being those who interact with pro-vaccination experts, where the latter group being more prone to active engagement in conversation in the vaccination topic. 

\begin{figure}[h] 
  \centering
  \includegraphics[width=.95\linewidth]{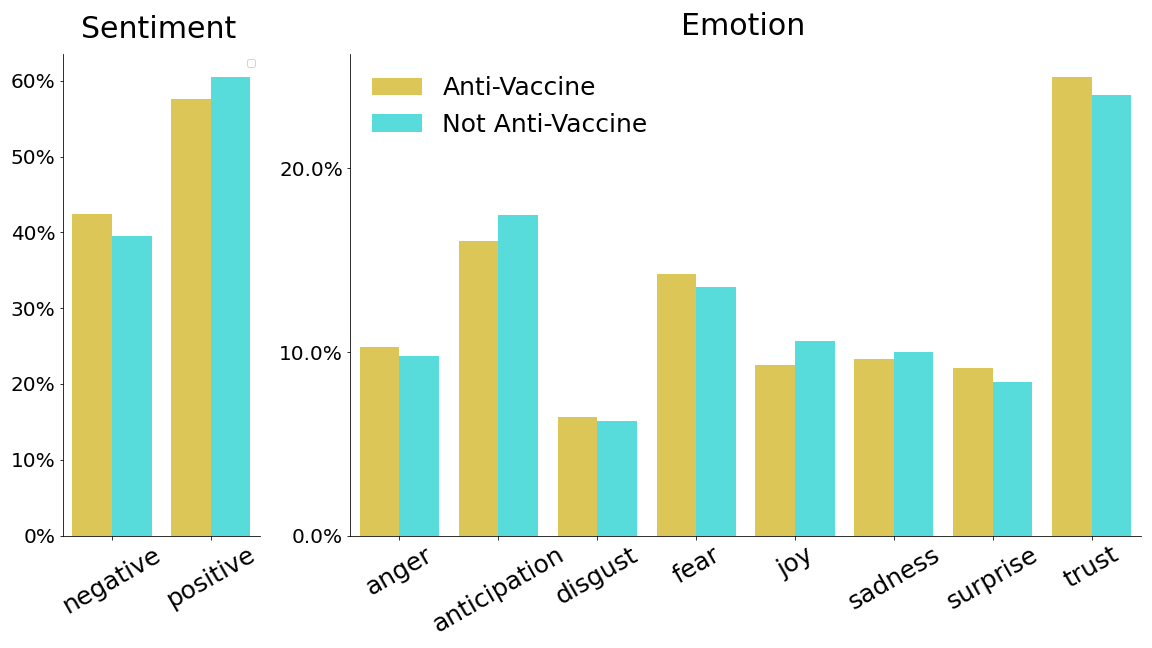}
  \caption{Emotion and Sentiment of tweets from Anti-Vaccination and Non-Anti-Vaccination accounts, based on NRC Lexicon~\cite{MohammadT13}.}
  \label{fig:affects}
\end{figure}

Using the NRC Lexicon~\cite{MohammadT13}, a sentiment and emotion lexicon, the average sentiment and emotion of tweets published by each class is displayed in Fig~\ref{fig:affects}. The anti-vaccination users lean towards a negative emotion, including displaying greater anger, disgust, fear, surprise, and trust. These users simultaneously have slightly lower sadness, anticipation, joy, and lower positive sentiment. The Mann-Whitney U Test revealed all features for both emotions and sentiments to have a statistically significant between-class difference in their distributions (p-value $<10^{-7}$).

\begin{figure*}[h] 
\centering
 \includegraphics[width=.8\linewidth]{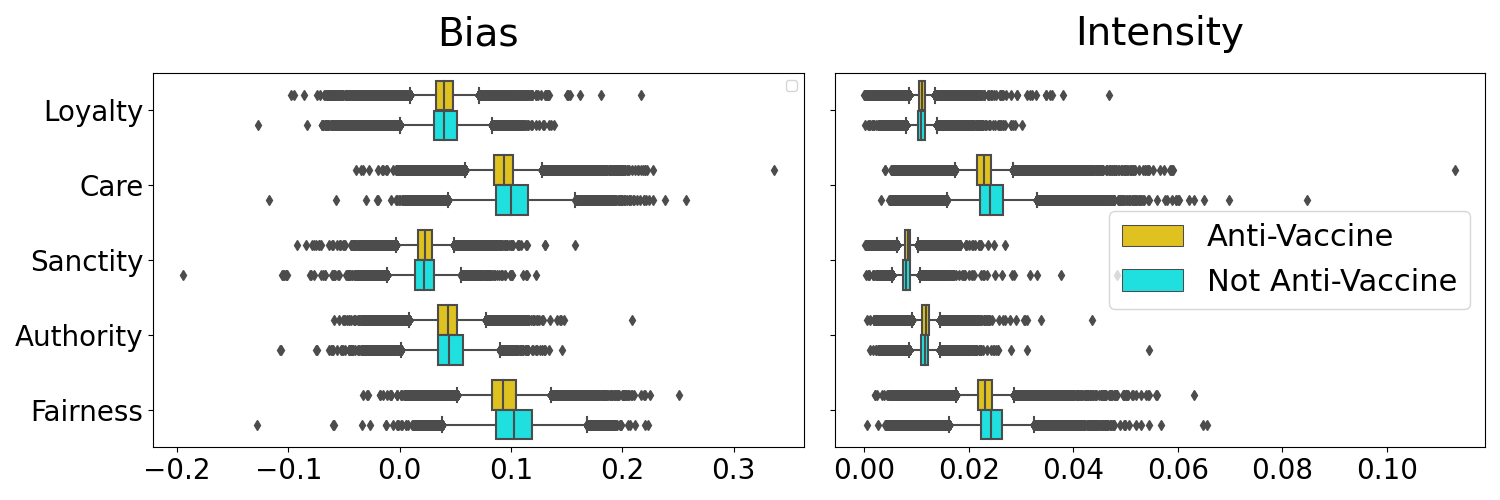}
\caption{Moral Foundations of anti-vaccine and regular users \cite{MoralFoundations}. (a) Bias (the tendency of words to promote positive or negative aspects of each moral foundation) are typically less positive in anti-vaccine users. (b) Intensity (the focus of words towards particular moral foundations) is also typically lower for anti-vaccine users.}
    \label{fig:MF}
\end{figure*}

An analysis of the Moral Framing associated with each group, based on the Modal Foundations Theory, was performed leveraging the Moral Foundations FrameAxis~\cite{Mokhberian2020} and is illustrated in Figure~\ref{fig:MF}. Moral foundation theory considers five basic moral foundations: loyalty, care, sanctity, authority, and fairness, used across cultures to determine morality~\cite{MoralFoundations}. Determining the morals in text is difficult, but one method to address this is by using word embeddings~\cite{Mokhberian2020}, and based on FrameAxis~\cite{FrameAxis}. 

We analyze two metrics from this method called bias and intensity. Bias tells us whether words tend to be associated with a positive or negative aspect of a moral dimension. For example, a highly positive loyalty bias means that a user is using words that are likely associated with being loyal rather than rebellious. Intensity tells us how prominently a particular moral dimension is used. A low intensity suggests users are not strongly associated with a particular moral dimension while high intensity suggests users are strongly associated (either positively or negatively) with a moral dimension. In more detail, the bias of a text towards each moral foundation axis is the weighted mean of the cosine similarity of the text's words with each axis. The absolute value captures the document's relevance to a moral dimension, while the positive sign denotes bias towards the Axis' positive pole of the Axis and negative sign denotes the opposite \cite{Mokhberian2020}. For example, a bias of -1 would indicate that a text is completely geared towards the negative aspect of a given dimension and a bias of +1 indicates the inclinations towards the positive aspect of a given dimension. Intensity denotes how frequently each moral dimension appears in the document with respect to the background distribution. Intensity disregards polarization, such that in situations where a text has both positive and negative bias, the terms cancel out each other, and the document does not display a significant bias towards any pole of that axis, yet intensity will show the relevance to that axis \cite{Mokhberian2020}. Typically most words spoken within a tweet are not moral in nature, and therefore most tweets have values near zero. 

%\begin{table}[h]
%\centering
%\caption{Median bias}
%\label{tab:MFFA-Bias}
%\small
%\begin{tabular}{r c c} 
% \toprule
% \textbf{Metric} & \textbf{Anti-Vaccine Score} & \textbf{Not Anti-Vaccine Score}  \\
% Authority & 0.0388 & 0.0542 \\ 
% Care & 0.0894 & 0.1062  \\
% Fairness  & 0.0913 & 0.1115  \\
% Loyalty  & 0.0391 & 0.0494  \\
% Sanctity  & 0.0232 & 0.0306  \\
%\bottomrule
%\end{tabular}
%\end{table}

%\begin{table}[h]
%\centering
%\caption{Median intensity}
%\label{tab:MFFA-Intensity}
%\small
%\begin{tabular}{r c c} 
% \toprule
% \textbf{Metric} & \textbf{Anti-Vaccine Score} & \textbf{Not Anti-Vaccine Score}  \\
% \midrule
% Authority & 0.0093 & 0.0101 \\ 
% Care & 0.0193 & 0.0217  \\
% Fairness  & 0.0201 & 0.0228  \\
% Loyalty  & 0.0088 & 0.0096  \\
% Sanctity  & 0.0063 & 0.0063  \\
%\bottomrule
%\end{tabular}
%\end{table}

The median values of each metric for anti-vaccine and regular users are statistically significant different (Mann-Whitney U Test p-value $<10^{-6}$). More specifically, we find that anti-vaccine users have a lower positive bias and intensity in loyalty, care, authority, and fairness, but slightly higher bias, and similarly more intense sanctity dimension (Figure~\ref{fig:MF}). This overall points to a lower focus by anti-vaccine users on positive morals, and less focus on most morals, which provides some support for anti-vaccine users being more anti-authority and anti-loyalty, and focusing less on care, while focusing on sanctity, which are morals like ''purity,'' ''immaculate,'' and ''clean.'' Notably, however, these differences are not very large, and despite a statistically significant difference in means, the distributions themselves are highly overlapped. Therefore, we should also keep in mind that people across this divide are more alike than different with regards to the morals they express.

\section{Discussion}

Overall, {AVAXTAR} delivers a fast and accurate open-source method to predicts users' vaccination attitudes, providing a foundational tool which facilitates further research in the field. By exploring the training dataset, we find anti-vaccine users are more negative and angrier and show a greater focus on politics (with common words like ``ballot'' and ``trump''). Finally, the analysis of user morals, based on Moral Foundation Theory \cite{MoralFoundations}, demonstrates less positive pro-authority or pro-loyalty bias, and a greater focus on sanctity and purity, perhaps because anti-vaccine users view vaccination as impure. This agrees with earlier work by Amin et al.~(\citeyear{amin2017association}), who found vaccine hesitant groups strongly emphasize purity, which suggests that the same moral framing has been used long before discussions of COVID-19 vaccine hesitancy.

\subsection{Limitations}
The present algorithm utilizes a dataset by Muric et al.~(\citeyear{Muric2021}) as anti-vaccine user samples. These data consist of automatically annotated anti-vaccine labels on Twitter accounts, via the usage of hashtags and keywords on published tweets. The not anti-vaccine samples includes both keyword- and hashtags-based data as well as users identified by their retweeting of prominent medical professionals. This method may generate both false positives and false negatives, which would then subsequently impact the accuracy of any model trained on the mislabeled data points. False positives can occur when a user does not display anti-vaccine sentiment, but made a publication including one of its associated hashtags or keywords, which could be due to typos, irony, or other reasons. False negatives can occur when a user clearly displays anti-vaccination sentiment, but happens not to use any of the hashtags or keywords employed in filtering for vaccine hesitant accounts. Since that same set of hashtags or keywords is used as negative filter in the non anti-vaccine class, such a user could end up being incorrectly included as a ``not anti-vaccine'' user. 

\subsection{Ethical Considerations}
The AVAXTAR system has the potential to positively impact our understanding of what leads to vaccine hesitancy. Conversely, the same ability of identifying anti-vaccine users could be leveraged to more perverse goals such as discrimination or targeted attacks. It is a socially tenuous line between incentivizing the population based on the vast research backing up vaccination and coercing or excluding persons given their stance on vaccination. Any efforts to define an ``in-group'' and an ``out-group'' can run the risk of further social division. For this reason, we make no judgement on whether a user with a probability score of, e.g., 0.6 or 0.99, is anti-vaccine or not. These probabilities should only be taken at face value. Lastly, we recognize that machine learning models can sometimes make wildly incorrect predictions, especially for edge cases. It is necessary to devise guardrails when building applications on top of AVAXTAR. 

The main benefit of AVAXTAR is that it enables future scientific research as well as development of applied solutions to the current challenge of vaccine hesitancy. Maximizing this positive impact demands free and open access to AVAXTAR. Such a publishing approach has to be weighted against the costs of possibly enabling the more perverse usages of our work. After significant consideration, we regard the risks of AVAXTAR being used to single out individuals to an extend that would be materially harmful to them as very unlikely. Moreover, if such aims are present, the malicious actor, who is supposedly willing to expend significant resources in its attack, can always replicate a similar system \cite{Yuan2019,carrieri2021predicting,huang2017examining,lincoln2022taking,Wang2020}. Conversely, we see the muffling of AVAXTAR technology as a hindrance to an entire community working towards ameliorating the side-effects of social media, such that limiting AVAXTAR's availability should cause more harm via preventing development on this area than it would prevent via curbing potential malicious use. Thus we convene that a public and freely distributed AVAXTAR is the preferable choice with regards to ethics.

\subsection{Future Work}
Correctly identifying which social media users propagate anti-vaccination sentiment is one of the many necessary steps to halt the current misinformation surge. It is therefore important to devise a science-based information campaign that targets vaccine-hesitant users, with the goal of halting the spread of misinformation. We also suggest an in-depth exploration of the extent to which Twitter-based models might rely on hashtags when generating predictions. This could take the form of a ablation-style study where models are compared against both full tweets as well as tweets stripped from their hashtags. The relationship between the ratio of replies to retweets or replies to likes might also provide an interesting avenue to exploration as it can signal posts which are highly inflammatory, be it with regards to vaccination or any other topic \cite{Minot2021}. Another important area of research is to predict which users are susceptible to anti-vaccine misinformation. This could be accomplished using data on the social media content a given user is viewing and interacting with, along with the existing data on user posts, as these combined data would allow researchers to understand what media consumption habits predicate a user joining the anti-vaccine movement.

\section*{Acknowledgements}
Funding for this work is provided through the ISI Exploratory Research Award.
% Anonymized for review purposes.

\section*{Conflicts of Interest}
The authors declare no conflicts of interest.

\bibliography{references.bib}
\end{document}